\documentclass[%
 aps,
 jmp,%
 amsmath,amssymb,
 reprint,%
superscriptaddress%
]{revtex4-2}

\usepackage{graphicx}
\usepackage{dcolumn}
\usepackage{bm}
\pdfoutput=1 

\begin{document}

\preprint{APS/123-QED}

\title{Interface stabilization in adhesion caused by elastohydrodynamic deformation}

\author{Preetika Karnal}
 \affiliation{Department of Chemical and Biomolecular Engineering, Johns Hopkins University, 3400 North Charles Street, Baltimore, Maryland 21218, United States}
 \affiliation{Department of Chemical and Biomolecular Engineering, Lehigh University, 124 E Morton St, Building 205, Bethlehem, Pennsylvania 18015, United States}
\author{Yumo Wang}%
 \affiliation{College of Mechanical and Transportation Engineering, China University of Petroleum, Beijing 102249, China}
\author{Anushka Jha}
 \affiliation{Department of Chemical and Biomolecular Engineering, Johns Hopkins University, 3400 North Charles Street, Baltimore, Maryland 21218, United States}%
\author{Stefan Gryska}
 \affiliation{3M Center, 3M Company, Building 201-4N-01, St. Paul, Minnesota 55144-1000, United States}
\author{Carlos Barrios}
 \affiliation{Adaptive3D, 608 Development Dr., Plano, Texas, 75074, United States}
\author{Joelle Frechette}
 \email[Corresponding author:]{jfrechette@berkeley.edu}
 \affiliation{Department of Chemical and Biomolecular Engineering, University of California Berkeley, California 94720, United States}



\begin{abstract}
Interfacial instabilities are common phenomena observed during adhesion measurements involving viscoelastic polymers or fluids. Typical probe-tack adhesion measurements with soft adhesives are conducted with rigid probes. However, in many settings, such as for medical applications, adhesives make and break contact from soft surfaces such as skin. Here we study how detachment from soft probes alters the debonding mechanism of a model viscoelastic polymer film. We demonstrate that detachment from a soft probe suppresses Saffman-Taylor instabilities commonly encountered in adhesion. We suggest the mechanism for interface stabilization is elastohydrodynamic deformation of the probe and propose a scaling for the onset of stabilization. 
\end{abstract}

\maketitle

 There is a wide interest in controlling interfacial instabilities, as they often affect the process in which they are formed.\cite{RN6, RN5, RN12, RN9, RN8, RN2, RN13, RN11, RN4, RN3, RN1} Interfacial instabilities can be a safety hazard for batteries,\cite{RN15, RN14} limit oil recovery,\cite{RN16} impact properties of graphene sheets,\cite{RN17} enhance the mixing of fluids,\cite{RN19, RN18} or guide the fabrication of soft materials\cite{RN20,RN21,RN22}. A common interfacial instability is the Saffman-Taylor type, manifested as undulating patterns formed in narrow gaps at fluid-fluid interfaces when a lower viscosity fluid displaces a higher viscosity fluid.\cite{RN28, RN26, RN27, RN25, RN23} Their onset can be controlled through low flow rates\cite{RN23} or local geometry \cite{RN5, RN6, RN32}. For example, elastic deformation of a membrane ahead of the fluid-fluid front alters the flow and suppress viscous instabilities.\cite{RN33, RN3} Due to their sensitivity to the flow profile, interfacial instabilities could potentially be manipulated in contact problems, such as in adhesion, where they are a source of energy dissipation.\cite{RN26, RN34,RN35, RN36, RN25} \par

\begin{figure} [ht]
\includegraphics[width=0.95 \columnwidth]{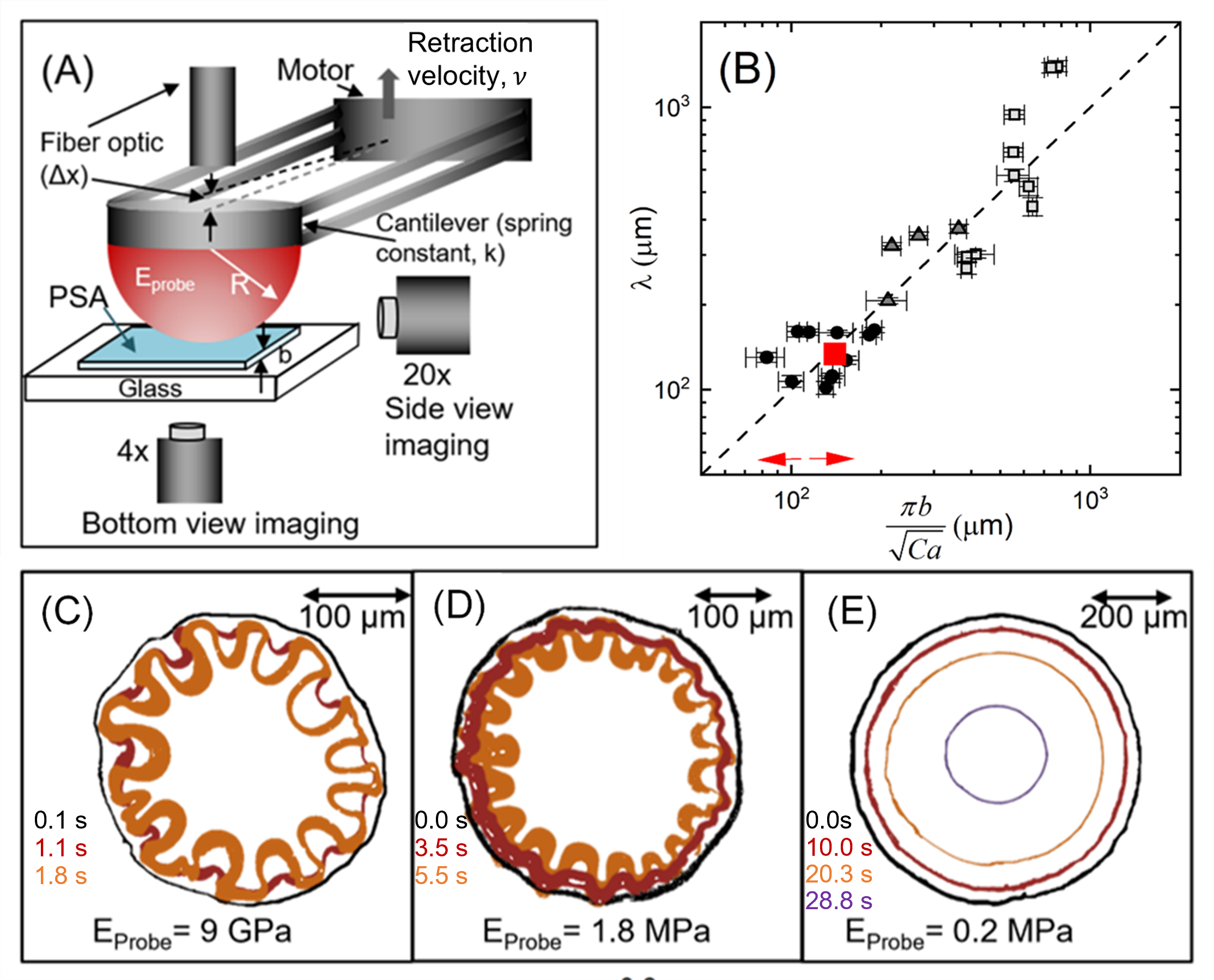}
\caption{(A) Schematic (not to scale) of measurements \cite{RN56}. (B) Validation of \textbf{Eqn. (1)}. Data from debonding between rigid probes and adhesive films of $b$= 25 $\mu$m (circle), 50 $\mu$m (triangle), and 100 $\mu$m (square). Arrows indicate the expected range for soft probes with $b$ = 25 $\mu$m. Unstable interface for $b=50$ $\mu$m for $E_\text{probe}$ =1.8 MPa (red square). (C-E) Superimposed contours of bottom view of the adhesive-air interface at different times during retraction from probes of different moduli. Initial contact line is in black. Interface moves radially inwards during retraction. Unstable interface (C,D) shows fingering as opposed to a stable interface (E).}
\end{figure}
Adhesion between two soft materials is ubiquitous during contact with skin with medical adhesives or flexible electronics.\cite{ RN42, RN41, RN39, RN43, RN40, RN74} Despite its technological significance, studies of adhesion between two soft materials are limited, but reveal qualitative differences from debonding from a rigid surface.\cite{RN47, RN49, RN51, RN50, RN48, RN46, RN52, RN54}. Here we show how the mode of debonding between a soft spherical probe and a thin viscoleastic adhesive film is altered as the compliance of the probe increases (\textbf{Fig. 1A}).  Saffman-Taylor instabilities are present during the detachment of a spherical rigid probe from a viscoelastic film (a variation of lifted Hele-Shaw cells). However, we find that the interface stabilizes when the compliance of the probe increases. The spherical probes studied are silicone elastomers for which we systematically vary the compliance, while the opposing surface is a soft viscoelastic pressure sensitive adhesive (PSA) film. We hypothesize that elastohydrodynamic deformation (EHD) of the spherical probe, caused by the viscous forces within the adhesive film during retraction, stabilizes the interface.\par

As control experiments, we measured the force in air during the detachment of rigid glass probes from the viscoelastic adhesive (thickness of $b= 25$ $\mu$m, Young’s modulus $\sim$30 kPa, \textbf{Fig. 1A}). The adhesion measurements are conducted on a microscope with bottom and side view imaging.\cite{RN55} During detachment, the adhesive-air interface is unstable and fingers form and grow until complete debonding (\textbf{Fig. 1C}). A distinguishing feature of interfacial instabilities in adhesion is the dependence of their wavelength, $\lambda$, on the detachment velocity. For a Saffman-Taylor instability $\lambda$ scales with the film thickness ($b$) and the Capillary number ($Ca = \eta^*U/\gamma$) as:
\begin{equation}\label{eqn:1}
    \lambda = \pi b/\sqrt{Ca} 
\end{equation}
where $\eta^*$ is the complex viscosity of the adhesive, $U$ is the radial velocity, and $\gamma$ is the surface tension of the adhesive-air interface.\cite{RN23, RN27, RN33, RN34} The complex viscosity accounts for the viscoelasticity of the adhesive. We measured adhesion for different detachment velocities and film thicknesses (25 -100 $\mu$m) and characterized fingering wavelengths at their onset (lowest strain in the films). We then compared our measurements to \textbf{Eqn. (1)} by determining the capillary number using the radial velocity of growing fingers’ apex, the complex viscosity $\eta^*$, and the surface tension (45 $\pm$ 2 mN/m).\cite{RN56, RN57} Agreement between data and \textbf{Eqn. (1)}, \textbf{Fig. 1B}, confirms the presence of Saffman-Taylor instabilities\cite{RN56}. In contrast, an elastic instability in the PSA would have a wavelength that only depends on the thickness of the adhesive ($\lambda_e=$4b)\cite{RN56, RN58, RN59, RN60, RN61}, and quadruple as we quadruple the film thickness. Instead, the change in wavelength if we quadruple the thickness increases by a factor of 3-12 dependending on the velocity, with the wavelength decreasing as the velocity increases, both characteristic of Saffman-Taylor instabilities.\par 

We then repeat the same measurements, but with silicone probes of increasing compliance. The compliant probes are made of PDMS (polydimethyl siloxane) of different crosslinking ratios that were extracted after curing to remove unreacted oligomers, and treated with plasma to render their surface hydrophilic. The soft probes have nearly identical geometry and surface energy as the rigid probes, but with a Young's modulus that varies from $\sim$2 MPa to $\sim$0.2 MPa.\cite{RN56}  We estimate the $Ca$ of the adhesive film during the detachment and found it comparable to values for rigid probes that displayed Saffman-Taylor instabilities (red arrows, \textbf{Fig. 1B}). Detachment with the stiffer PDMS leads to an unstable interface, but the interface stabilizes for softer probes (\textbf{Fig. 1D,E}).\par
\begin{figure}[!]
\centering
\includegraphics[width=0.95 \columnwidth]{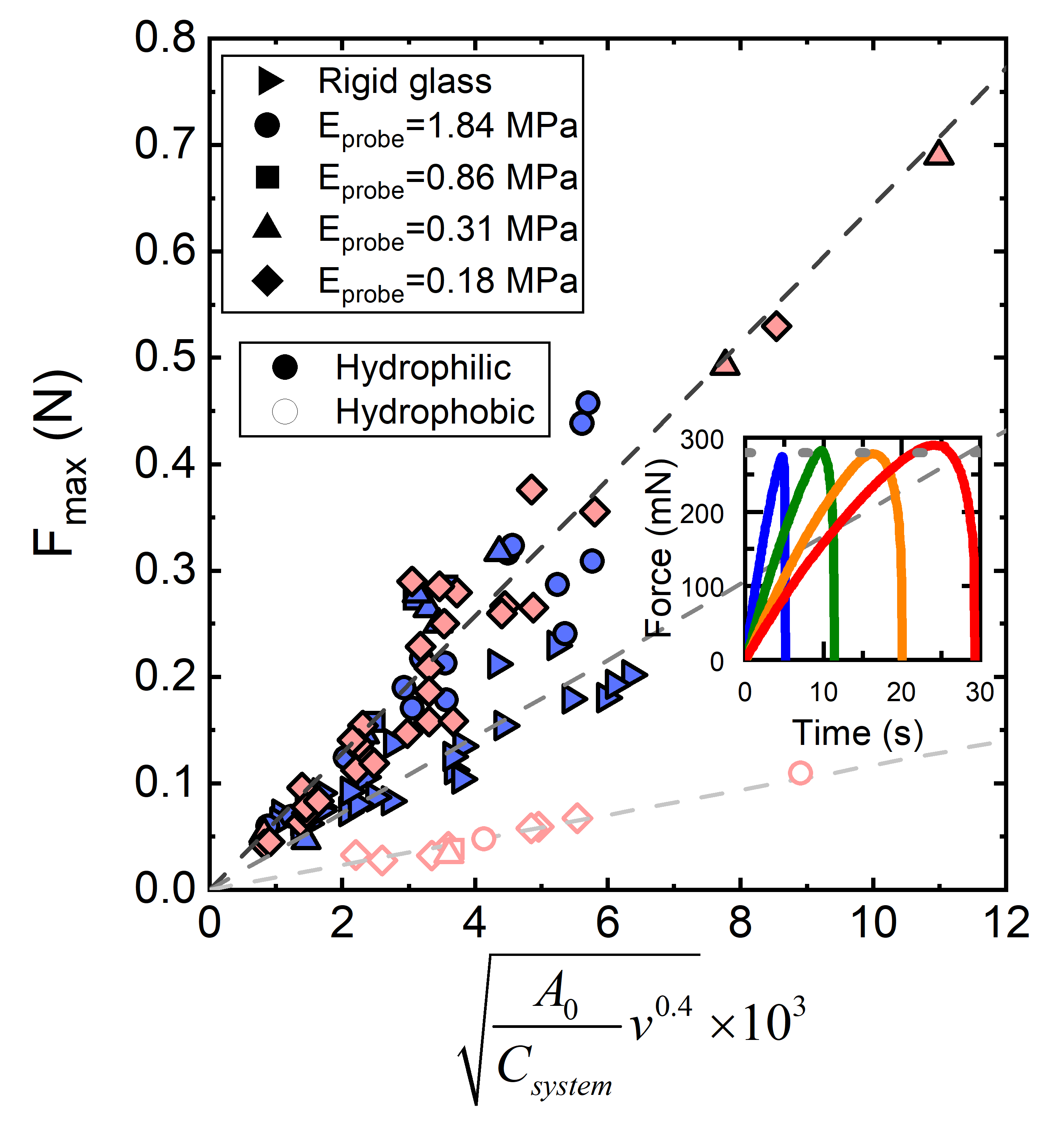}
\caption{Adhesive strength for different probes and retraction velocities. The slope $\sim\left(2\sqrt{G_0/v_\text{ref}^{0.4}}\right)$ increases with probe surface energy. There is no distinction in the adhesive strength for a stable (pink) or unstable (blue) interface. Inset: Debonding curve between soft PDMS probes and adhesive films at $v=50 \mu$m/s. Increase in probe compliance leads decreases the slope. The maximum force,  $F_\text{max}$, is independent of probe modulus.   }
\end{figure}
\begin{figure*}[!]
\centering
\includegraphics[width=0.85\textwidth]{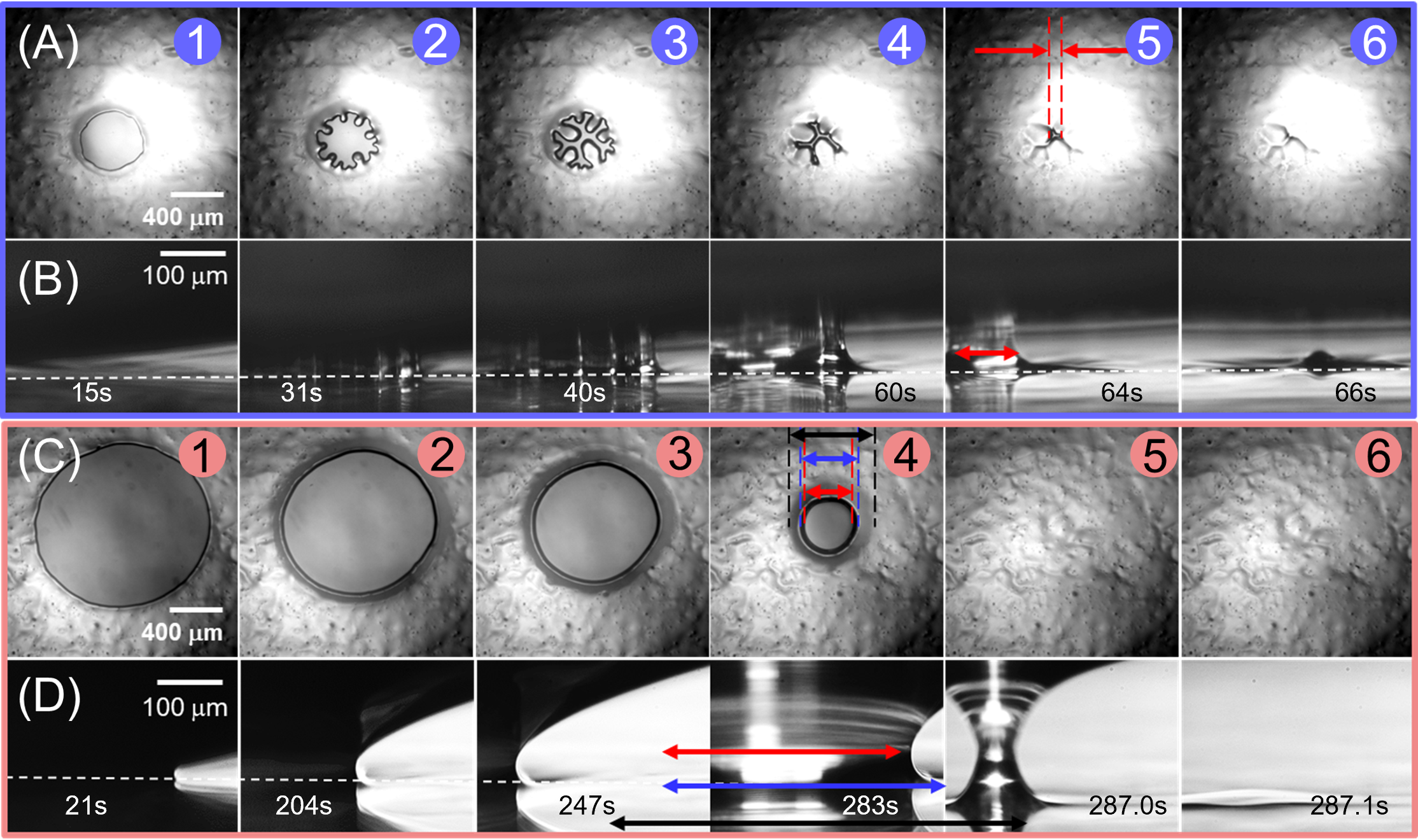}
\caption{Side and bottom view images during debonding over time. Rigid probe (A) bottom and (B) side views. Soft probe, $E_\text{probe}$ = 0.18 MPa (C) bottom and (D) side views. Instabilities are present during debonding from the rigid probe. Side view images (B) show stretching of adhesive. For the soft probe (C) the interface is stable, side views (D) show probe deformation. Note the different scale and magnification between the side and bottom views; the arrows represent the same dimension.}
\end{figure*}
Due to confinement, the compliance of the adhesive film is smaller than its bulk counterpart, and also smaller than all soft probes investigated.\cite{RN56} While a PSA is a viscoelastic solid, a simple stress-strain model where the thin adhesive film is in series with a soft probe ($k_\text{PSA} \gg k_\text{Probe}$) suggests a significant dissipative response due to the complex viscosity of the adhesive.\cite{RN56} Therefore, even if the adhesive film is a solid its dynamic response is dominated by viscoelasticity.  Moreover, recent work shows that in the case of elastic instabilities the interface can become stable as the probe modulus increases, the opposite of our observations.\cite{RN73} .\par
As the probe compliance increases, the interface becomes stable during detachment (\textbf{Fig. 1C-1E}). Because only the compliance of the probe is varied (and not its surface energy), the experiments suggest the importance of compliance on interface stabilization.\cite{RN56} The transition to a stable interface also has no impact on the adhesive strength ($F_\text{max}$  in \textbf{Fig. 2 Inset }). For the same debonding velocity the adhesive strength is nearly the same for all probe moduli, without any distinction between stable and unstable interfaces. For the sphere-plane geometry the adhesive strength is independent of compliance, but the mode of failure can affect the force profile.\cite{RN63, RN64}  A small plateau in force was also observed with the onset of fingering instabilities when lifting rigid plates confining viscous fluid,\cite{RN35}, whereas adhesion-induced elastic instabilities increased the resistance to deformation leading to higher forces.\cite{RN62}

We also find that stabilization of the interface is not due to a change in probe surface energy. The relationship between the adhesive strength ($F_\text{max}$), debonding velocity ($v$), and compliance is well-established and given by:\par
\begin{equation}\label{eqn:2}
    F_\text{max}=2\left[\frac{A_0}{C_\text{sys}}G_0\left(\frac{v}{v_{ref}} \right)^n \right]^\frac{1}{2},
\end{equation}
where $G_0$ is the intrinsic strain energy release rate, $A_0$ is the maximum contact area, $C_\text{sys}$ is the system compliance, $v$ is the debonding velocity, and $n$ is an empirical constant, here $n=0.4$.\cite{RN63, RN64, RN66,RN56} Therefore, for a constant apparent work of adhesion we expect a linear relationship between $F_\text{max}$ and $\sqrt{A_0/C_\text{sys} v^{0.4}}$ with a slope $2\sqrt{G_0/v_\text{ref}^{0.4} }$. Adhesion follows well the established force scaling relationship, with no departure from the linear relationship that would indicate a change in surface energy for softer PDMS probes. Data for the hydrophilic PDMS includes the adhesive strength for probes with elastic moduli between 0.18 and 1.8 MPa (\textbf{Fig. 2}).\cite{RN56} The linear relationship observed across PDMS probe moduli confirms the constant apparent surface energy. This linear relationship also holds for probes of different surface energy, but with a different slope (silica and hydrophobic PDMS, \textbf{Fig. 2}).  \par

Side view imaging shows that transition to a stable interface is accompanied by significant elastic deformation of the probe, \textbf{Fig. 3}. The forces resisting the probe’s upward motion within the adhesive film cause elastic deformation of the probe and appear to be stabilizing the interface.  For hydrophilic PDMS probes, interface stabilization occurs despite having the same intrinsic surface energy.  Using $G_\text{0}/E_\text{eff}a$ we evaluate the intrinsic strain energy release rate normalized with the contact compliance (or the elastoadhesive length normalized with the contact radius)\cite{RN75}, where the effective modulus is $E_\text{eff}=3/4C_{sys}a$ , and $a$ is the contact radius. This quantity represents the ability of a material to resist crack propagation through elasticity.  Changes in the relative importance between contact compliance and surface energy in the contact region, $G_\text{0}a^{2}$, \textbf{Fig. 4A}, do not delineate stable from unstable interfaces.  In other words, the deformation of the probe is not dominated by an increased contribution from the surface energy as the probe modulus decreases.\par 
Here, debonding occurs between a soft probe and a viscoelastic adhesive. At any given time the measured force is due to surface, viscoelastic, and elastic (probe deformation) contributions. The elastic (probe deformation) and viscous (“flow” of the adhesive) forces are highly coupled. Elastohydrodynamic deformation occurs when the viscous forces in a fluid are strong enough to cause elastic deformation to an opposing surfaces.\cite{RN67, RN68, RN71, RN72} We hypothesize that the probe deformation alters the pressure distribution within the adhesive film leading to a suppression of Saffman-Taylor instabilities.  The relative importance of elastohydrodynamic deformation can be estimated through an elasticity parameter, $\epsilon$ (\textbf{Eqn. (3)}), obtained from non-dimensionalization of the lubrication equation:	\cite{RN67, RN68, RN69}
\begin{equation}\label{eqn:3}
    \epsilon = \frac{\eta^*v R^{1.5}}{E_\text{probe}^* b^{2.5}}.
\end{equation}
The elasticity parameter can be viewed as a ratio between elastic forces within the probe and viscous forces within the adhesive film. As $\epsilon$ increases the elastic deformation of the probe ($w$) increases,. For low $\epsilon$ viscous forces do not cause probe deformation. We previously found that the dimensionless central deformation ($\widehat{w} =w/b$) of a spherical probe scales with $(6\epsilon)^{0.4}$.\cite{RN70} 

\begin{figure}[!]
\centering
\includegraphics[width=0.95 \columnwidth]{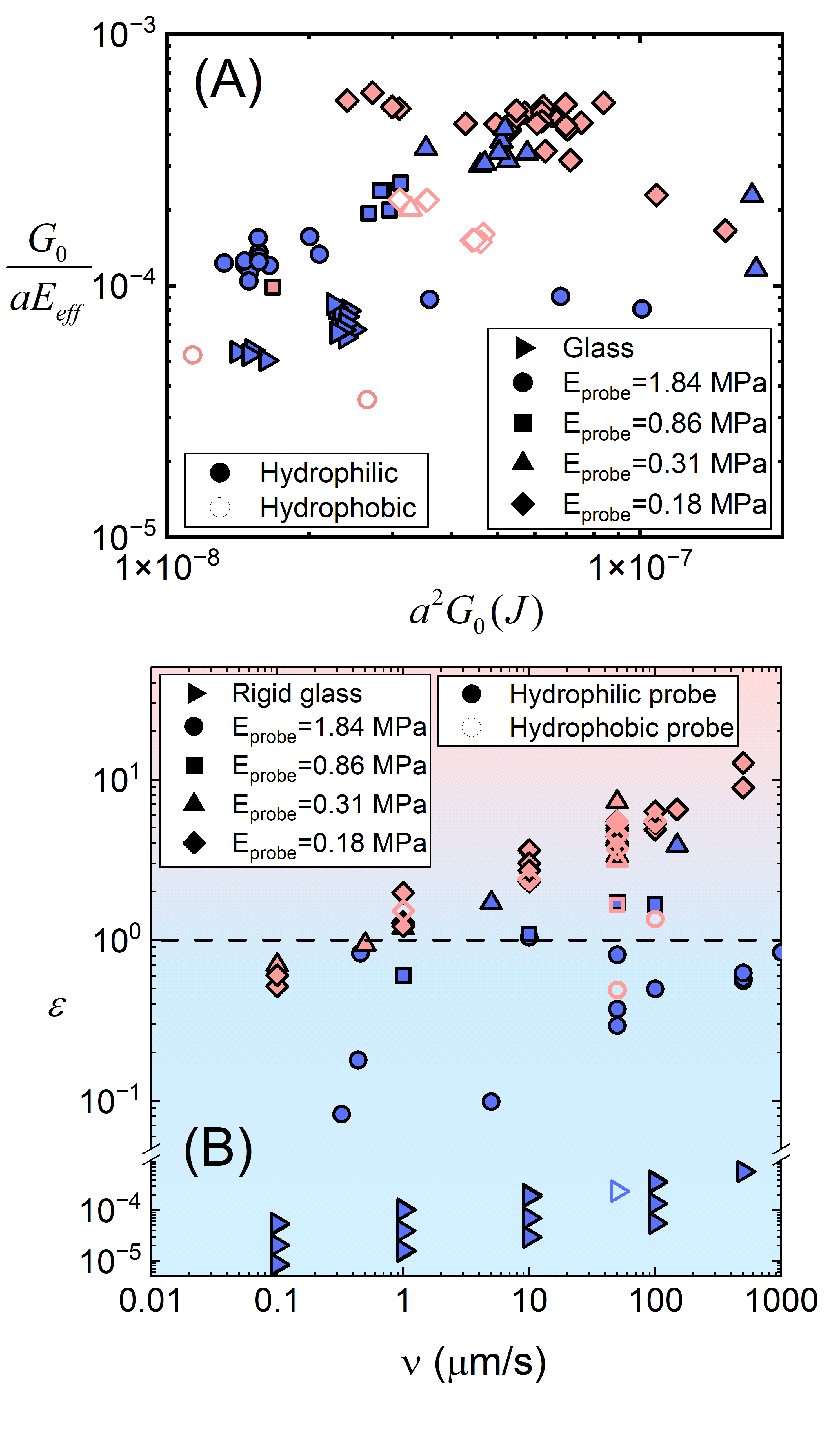}
\caption{(A) Elastoadhesive length normalized by the contact radius  vs effective surface energy for all probes. (B) Elasticity parameter ($\epsilon$) vs debonding velocity  (${v}$) . The transition from unstable to stable interface is observed around  $\epsilon=1$ (black dotted line). Data includes adhesive with $b=$ 25, $b=$ 50, $b=$ 100 $\mu$m and $R$ between 4.5 mm – 14 mm and shows unstable interface (blue) and stable interface (pink).}
\end{figure}

\begin{figure}[!]
\centering
\includegraphics[width=0.95 \columnwidth]{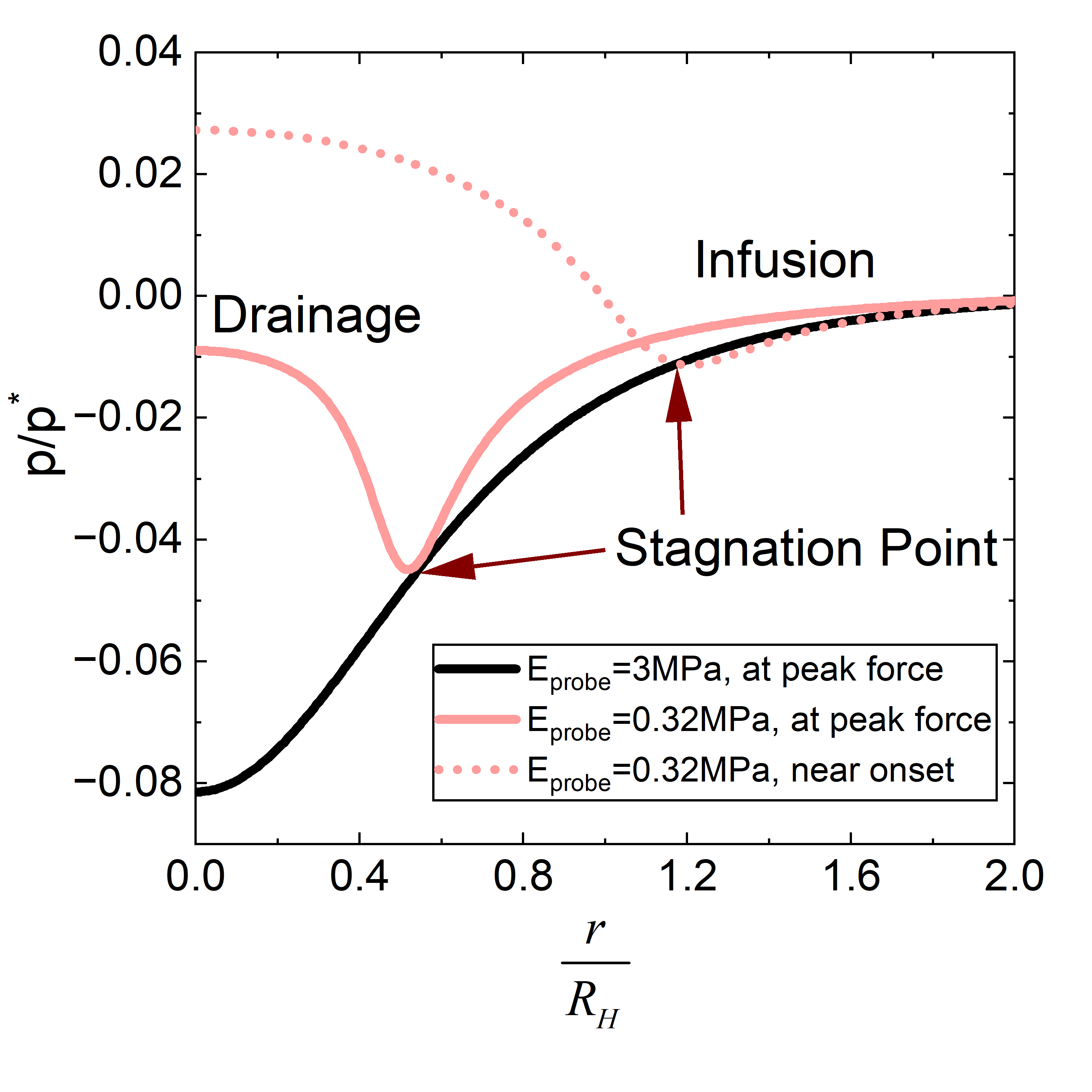}
\caption{ Dimensionless pressure ($p^*=\eta vR/b^2$) versus dimensionless radial position ($R_H= \sqrt{2Rb}$) obtained from modeling the detachment of soft  ($E_\text{probe}$=0.32 MPa) and stiff ($E_\text{probe}$=3 MPa) PDMS probes of R = 6 mm at 50 $\mu$m/s and $\eta$=1000 Pa.s, b=20 $\mu$m. Retraction of the soft probe leads to lower fluid pressure and appearance of stagnation point delineating drainage and infusion regions.}
\end{figure}

A plot of $\epsilon$ as a function of  debonding velocity ($v$) shows a clear demarcation between stable and unstable interfaces (\textbf{Fig. 4B}). The transition to a stable interface occurs across different materials systems and experimental parameters: probe modulus, radius, detachment velocity, and film thickness. The transition between an unstable and stable interface occurs around $\epsilon=1$, when the elastic forces in the probe begin to dominate over the viscous forces in the adhesive film. The transition to a stable interface as the elasticity parameter increases supports the hypothesis that elastohydrodynamic deformation of the probes suppress the fingering instabilities. \par
As the velocity increases the adhesive strength increases, and stabilization of the interface shifts to higher $\epsilon$ (\textbf{Fig. 4B}). We compare the role of debonding velocity on the probe deformation and the pressure within the film. An increase in $v$ will increase the pressure within the adhesive film, which has a destabilizing tendency for the interface. However, increasing the velocity also increases the probe deformation, which we hypothesize stabilizes the interface. Non-dimensionalization of the lubrication equation leads to a characteristic pressure in the fluid, $p^*=\eta vR/b^2$.\cite{RN70} For a viscoelastic film,  $p^*=\eta^* vR/b^2$, therefore $p^*\propto v^{(1-m)}$ with the dependence of the complex viscosity on velocity. For our material here $m$=0.725, giving $p^*\propto v^{0.28}$. Moreover, the dimensionless central probe deformation scales as $\hat{w}\sim v^{0.4(1-m)}$, and specifically for our material system $\hat{w}\sim v^{0.11}$. Therefore, as $v$ increases the pressure within the film (${p^*} \sim v^{0.28}$) increases faster than the deformation of the probe ($\widehat{w} \sim v^{0.11}$). The faster increase in pressure within the film as $v$ increases would necessitate larger probe deformations to stabilize the interface, thus a higher elasticity parameters is needed for stabilization.\par
We study the relationship between elastohydrodynamic deformation and adhesive film pressure by modeling debonding between a soft probe and a rigid surface submerged in a Newtonian fluid.\cite{RN56} In the model the fluid viscosity is comparable to the complex viscosity of the adhesive. This model is a highly simplified version of our experiments, in that the adhesive is treated as a viscous fluid without an air-adhesive interface present. We extract the pressure profile during detachment for both rigid and soft probes and obtain lower fluid pressure with the soft probe, \textbf{Fig. 5}, which would have a stabilizing effect. We also observe that the elastohydrodynamic probe deformation leads to a non-monotonous pressure drop within the fluid. In contrast, the pressure distribution is monotonic during the detachment from a rigid probe. Moreover, deformation of the soft probe leads to a negative pressure gradient at the center point, causing the fluid \textit{drainage} from the center during detachment, while further away from the center the pressure drop is positive leading to the expected fluid \textit{infusion}. Between the drainage and infusion regions there is a stagnation point where the pressure gradient is zero. The stagnation point moves towards the center of the probe during retraction (\textbf{Fig. 5}). Because of incompressibility, the surfaces initially move closer at the center point during detachment. The combination of lower pressure and a stagnation point could suppress the Saffman-Taylor instabilities during the detachment from a soft probe, and will be the subject of future studies.

In summary, the detachment of a viscoelastic adhesive from soft surfaces suppresses the onset of Saffman-Taylor instabilities. While elasticity has been shown previously to impact Saffman-Taylor instabilties, we show here the connection with adhesion. Controlling the mode of failure during debonding between soft materials and could impact adhesion (and pain) with skin. We attribute stabilization of the interface to elastohydrodynamic deformation of the probe caused by viscoelasticity. The elasticity parameter can serve as a guide for interfacial stability. A simple model shows that replacing a rigid probe with a soft one leads to a decrease in the pressure drop and the appearance of a stagnation point within the film, both could lead to interface stabilization. Further studies are necessary to better understand the detachment process between two soft materials and the stabilization of the interface. 

\textit{Acknowledgements:} This work was supported by 3M and by the National Science
Foundation (NSF-CMMI 1728082). Y.W. also acknowledges support from the National Natural Science Foundation of China (Grant No. 51804319.)

\nocite{*}

\bibliography{main.bib}


\end{document}